\documentclass[%
 reprint,
 amsmath,amssymb,
 aps,
]{revtex4-1}

\usepackage{graphicx}
\usepackage{dcolumn}
\usepackage{bm}
\usepackage{hyperref}

\usepackage{subfigure}
\usepackage{multirow}

\begin{document}

\preprint{APS/123-QED}

\title{On the relation between active population and infection rate of COVID19}

\author{Takashi Shimada}
 \altaffiliation[Also at ]{Mathematics and Informatics Center, The University of Tokyo}
  \email{shimada@sys.t.u-tokyo.ac.jp}
\author{Yoshiyuki Suimon, Kiyoshi Izumi}%
\affiliation{%
Department of Systems Innovation, Graduate School of Engineering, The University of Tokyo
}%

\date{\today}

\begin{abstract}
The relation between the number of passengers in the main stations and the infection rate of COVID19 in Tokyo is empirically studied.
Our analysis based on conventional compartment model suggests:
1) Average time from the true day of infection to the day the infections are reported is about $15$ days.
2) The scaling relation between the density of active population and the infection rate suggests that the increase of infection rate is linear to the active population rather than quadratic, as that is assumed in the conventional SIR model.
3) Notable deviations from the overall scaling relation seems to correspond to the change of the peoples's behavior in response to the public announcements of action regulation.
\end{abstract}

\maketitle


\section{Introduction}
Reducing the number of people in public area by some means of regulations is one of the most major strategies to slow or prevent the spread of epidemics such as COVID-19.
The main drawback of this measure is its negative impact on our economical and social activity.
Therefore it is practically very important to quantify the effect of reduction of population density in public area on the reduction of infection rate.
It is also essential for wide field including ecology, epidemiology, and sociology to understand the functional relation between the population density and the infection (contact) rate in real case \cite{Murray2002, Murray2003, Begon1996}.

Here we report our empirical study on the relation between the number of passengers in mains stations of Tokyo metropolitan area and the increase in the number of infections.

\begin{figure}[bt]
\includegraphics[width=1.0\linewidth]{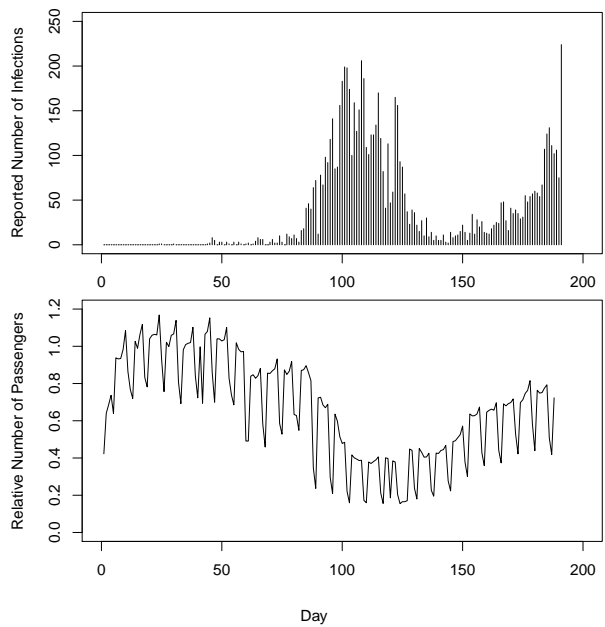}
\caption{
Daily data for infection rate and crowd in this study. 
(Top) Daily number of infection reports in Tokyo metropolitan area \cite{InfectionData}.
(Bottom) The daily total number of passengers in main stations of Tokyo metropolitan area (Shinjuku, Shibuya, Ikebukuro, Shinagawa, and Tokyo) detected by a mobile carrier company\cite{CrowdData, Suimon_report}, relative to its typical average.
}
\label{fig_InfectionPassengers}
\end{figure}

\section{Method}
\subsection{Calculating effective reproduction number}
Let us start from taking a simple compartmental model in the difference equation form:
\begin{eqnarray}
	\Delta S_t &=& S_{t+1} - S_t = -\beta C(S_t, I_t), \cr
	\Delta I_t &=& I_{t+1} -  I_{t} = \beta C(S_t, I_t) - \gamma I_t , \cr
	\Delta R_t &=& R_{t+1} - R_{t} = \gamma I_{t} ,
	\label{eq_SIR}
\end{eqnarray}
where $S_t, I_t$, and $R_t$ are the numbers of susceptible, infected, and recovered (removed) people at a discrete time (i.e. day in this study) $t$, respectively.
Positive constant parameters $\beta$ and $\gamma$ represent the (effective) infection rates per human contact and the recovery rate, respectively.
In the following we take $\gamma = 14.0^{-1}$ [day], i.e. the average time till recover is 2 weeks, and it will be confirmed later.
The function $C(S_t, I_t)$ denotes the rate of human contacts. For example, taking a Lotka-Volterra type contact rate: $C \propto S_t I_t$ corresponds to the SIR model.

As shown in Fig. \ref{fig_InfectionPassengers}, the typical data available is the daily report of newly infected people $O_{t+t_d} = -\Delta S_{t}$.
The delay time $t_d$ is the difference between the day of the report of an infection and the real moment of infection, which is typically said to be around 2 weeks, consists of the time from the day of infection to the day the person went to a hospital, the time the institution took on PCR-test, and the time spent to include that result to the report, etc. 
We will estimate $t_d$ from the data.

The effective reproduction number calculated as
\begin{equation}
	{\cal R}^e_t
	\equiv
	\frac{-\Delta S_{t}}{\Delta R_{t}}
	=
	\frac{O_{t+t_d}}{\displaystyle \gamma \left( \sum_{\tau=\tau_0}^{t} O_{\tau+t_d} \ {\rm e}^{-\gamma(t-\tau)} \right)}
	\label{eq_Re}
\end{equation}
is the key quantity to predict whether the epidemic process will grow (${\cal R}^e > 1$) or decay (${\cal R}^e < 1$) if the effect of in- and out-flows are negligible.

Note that, because $O_t$ is known to have strong systematic dependance on days of week which comes from the difference in the activity such as testing and reporting,
we take 7 day average for $R^e_t$ in the following analysis.

\subsection{Estimating the relative number of active people}
Another key data for our study is the change in the number of passengers in Tokyo main stations (Shinjuku, Shibuya, Ikebukuro, Shinagawa, and Tokyo), estimated from the mobile phone data (Fig. \ref{fig_InfectionPassengers}, Appendix).
We take the sum of observed passengers $N_t$ to simply estimate the number of people who were not staying home on that day $N^{active}_t$,
i.e. $N^{active}_t \propto N_t$.

\subsection{Relating ${\cal R}^e$ and $n_t$}
Let $n_t = N_t/\bar{N}$ be the number of people acting outside on a certain day relative to a normal typical average $\bar{N}$.
Then the number of people who had a chance to contact should be replaced by the actual active population
$\displaystyle \tilde{S_t} = n_t S_t$ and $\displaystyle \tilde{I_t} = n_t I_t.$
Therefore, assuming the SIR (Lotka-Volterra) type contact rate $C = C_0 S_t I_t$,
the rate of infection and the effective reproduction number should depend proportionally to the square of $n_t$
\begin{eqnarray}
	O_{t+t_d} &=& - \Delta S_t =  \left( \beta C_0 S_t I_t  \right) n_t^2, \\
	{\cal R}^e_t &=& \frac{O_{t+t_d}}{\gamma I_t} \approx \left( \frac{\beta C_0}{\gamma} \right) n_t^2.
\end{eqnarray}
In the second equation we assume that the accumulated number of infections is negligibly smaller than the total population,
i.e. $\displaystyle 1 - S_t/S_0 \approx 0$.
This assumption is strongly supported by the random antibody test conducted in early June in Tokyo area, which reported that only $0.10$\% of people have antibody
\cite{AntibodyTestJune2020}.
Any deviation from this simple scaling relation between $N_t$ and ${\cal R}^e$ implies that the simple SIR-model-type contact rate,
which models the chance of human contact based on the well-mixed picture, does not hold in reality.

\section{Results}

\begin{figure}[tb]
\includegraphics[width=1.0\linewidth]{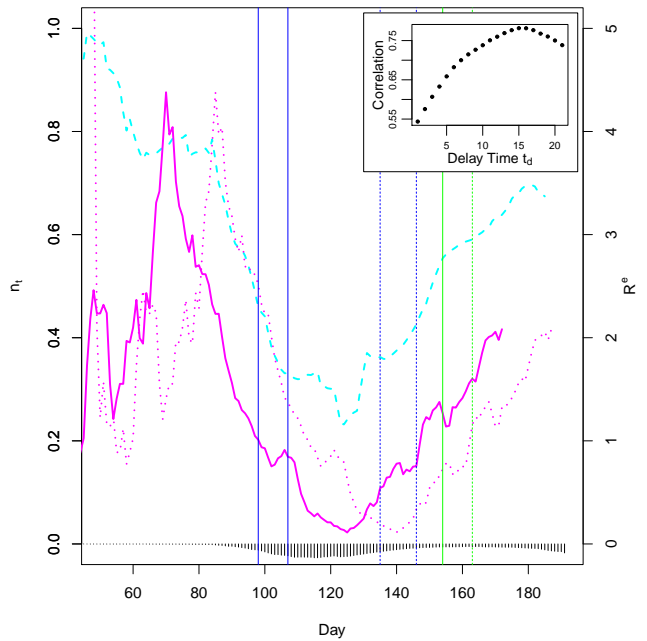}
\caption{
The time series of effective reproduction number $R^e_t$ for optimal $t_d = 15$ [day] (solid magenta line, axis to the left), with the number of passengers $N_t$ (dashed cyan line, axis to the right). Both time series are smoothed by taking 7 day average.
Dotted line is the $R^e$ with no consideration of delay time (i.e. $t_d = 0$) for comparison.
Inset shows the Spearman's rank correlation between $N$ and $R^e$ to given delay time $t_d$.
Vertical lines are the eye-guide for notable days of regulation (solid) or relaxation (dashed) in policy;
Day 98 (7th April): the first emergency declaration to limited area including Tokyo,
Day 107 (16th April): emergency declaration was expanded to entire country,
Day 135 (14th May): emergency declaration was canceled for the area except for Hokkaido, Saitama, Chiba, Tokyo, Kanagawa, Kyoto, Osaka, and Hyogo,
Day 146 (25th May): emergency declaration was canceled for entire country,
Day 154 (2nd June): declaration of ``Tokyo-alert'', and
Day 163 (11th June): cancelation of ``Tokyo-alert''.
The reported number of infections are also shown inversely for eye-guide. 
}
\label{fig:Conf_Tdelay}
\end{figure}

\begin{figure*}[tb]
\includegraphics[width=0.45\linewidth]{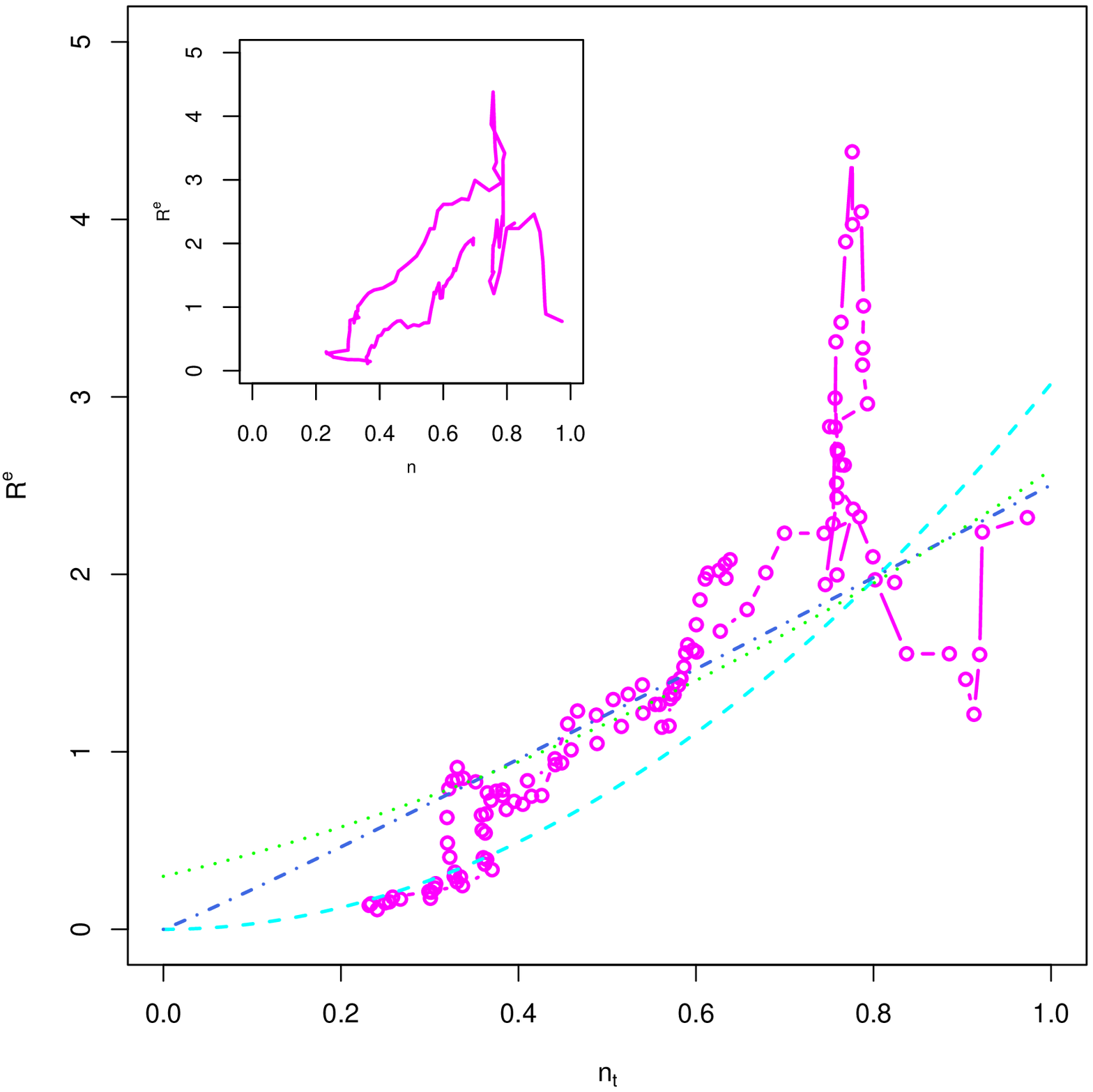}
\includegraphics[width=0.45\linewidth]{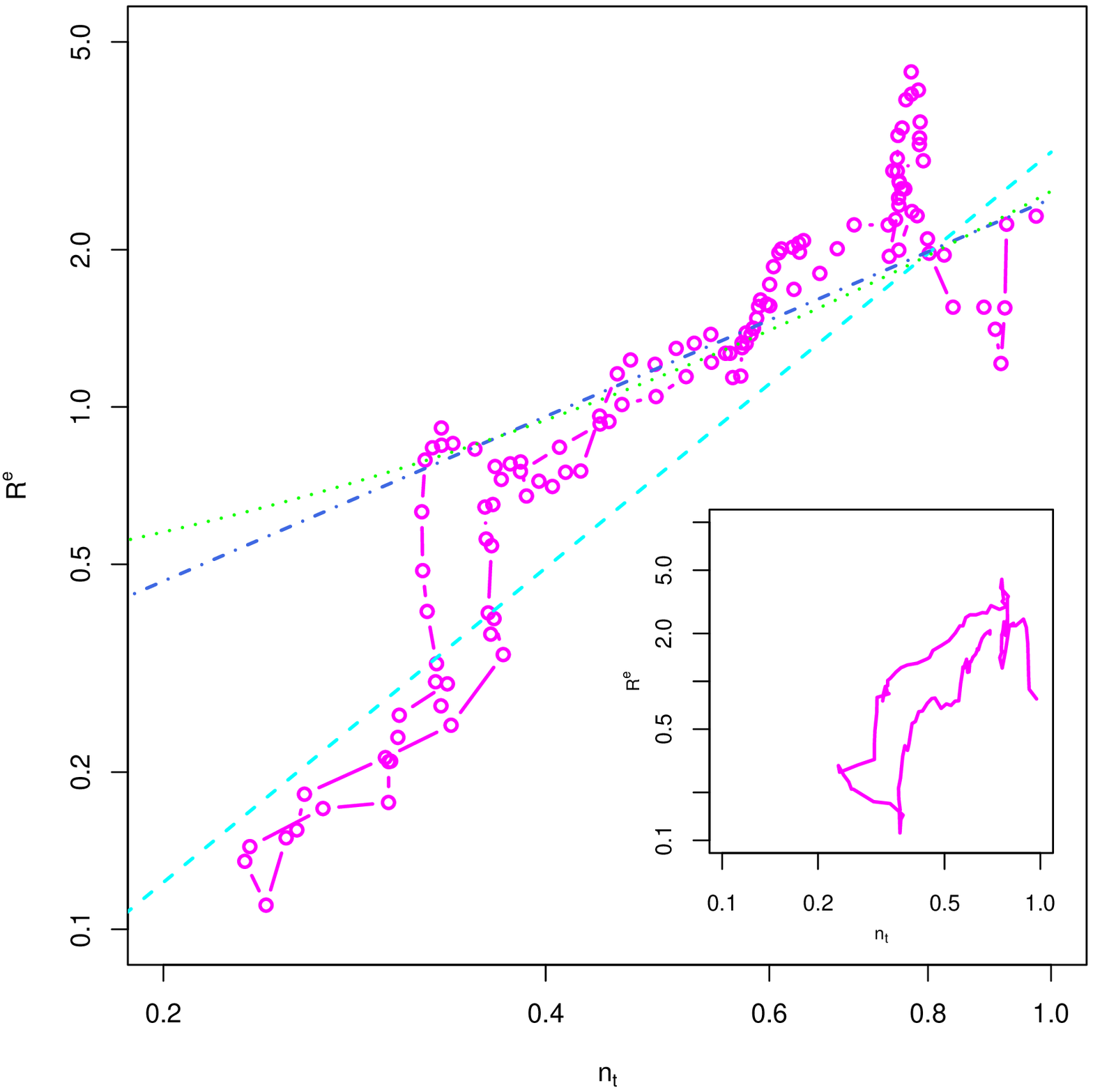}
\caption{
The scaling relation between the relative number of passengers and the effective reproduction number $R^e_t$, which is proportional to the rate of infectious contacts.
Lines show the fittings by
quadratic function: $R^{LV1}_t =R_0 n_t^2$ (cyan dashed line),
quadratic term with constant base population: $R^{LV2}_t = R_0 \left( \frac{n_t+n_0}{1+n_0} \right)^2$ (green dotted line),
and the power-law: $R^{DD}_t = R_0 n_t^{2-\alpha}$ (blue dot-dashed line).
Inset figures show the loop shape of the same $n_t-R^e_t$ plot with non-optimal time delay ($t_d = 7$ [day]).
}
\label{fig:fittings}
\end{figure*}

\subsection{Determining the typical delay time from the infection to the detection}
The dotted line in Fig. \ref{fig:Conf_Tdelay} shows the effective reproduction number calculated by eq. (\ref{eq_Re}), with no consideration about the delay time (i.e. $t_d = 0$).
The estimated value of ${\cal R}^e$ is in a plausible range $(< 3.0)$, except for the early period and the short period around the sharp peak.
The former is simply from the smallness of the denominator in eq. (\ref{eq_Re}) and the latter sharp peak period is considered to be from the inflow of infected people from abroad.

The unknown parameter that should to be determined first for our analysis is the average delay time $t_d$ from the actual day of infection to the day on which the infection is reported. As mentioned above, although it is generally said to be around 2 weeks in the case of Tokyo, we cannot determine this parameter solely from epidemiological knowledge.

To estimate this parameter, we take a natural model-free assumption:
{\it Infection rate is an increasing function of active population.}
Under this assumption, optimal $t_d$ should give the highest rank correlation between $N_t$ and $O_{t+t_d}$ (and hence ${\cal R}^e$).
From this criterion, the optimal delay time is found to be $t_d = 15$ [day] (both from Spearman's and Kendall's rank correlation coefficients, and for a certain range of $\gamma$. See Appendix for more detail). 
A good correlation between $N_t$ and $R^e_t$  (Speaman's rank correlation coefficient $r_s^* = 0.7815$) can be also easily confirmed in Fig. (\ref{fig:Conf_Tdelay}) by eye.

What is interesting, in addition to the overall correlation, is that the days of the changes of regulations policy on the people's activity coincide with the changes in $R^e_t$.
$R^e_t$ shows sharp drops soon after the days of first emergency declaration (to a part of Japan including Tokyo, 7th April) and its application to the entire country (16th April),
and also soon after the declaration of ``Tokyo-alert'' from the Tokyo metropolitan government.
Sharp rises are also found soon after the days of relaxing the policies. Those might be indications of the change in the people's behavior, rather than the change in the population outside.

\begin{table}[b]
	\begin{ruledtabular}
   	\begin{tabular}{l|c|c|c}
		Model & $R_0 $ & second parameter & deviance \\
		\hline &&&\\
		LV1: $\displaystyle {\cal R}_0 n_t^2$ & $3.045$ & - & 91.90  \\
		&&&\\
		LV2: $\displaystyle {\cal R}_0 \left( \frac{n_0+n_t}{n_0+1} \right)^2$ & $2.572$ & $n_0 = 0.5183$ & 71.43\\
		&&&\\
		DD: $\displaystyle \ {\cal R}_0 n_t^{2 - \alpha}$ & $2.489$ & $\alpha=0.9558$ & 66.37
	\end{tabular}
	\end{ruledtabular}
	\label{table:fittings}
\end{table}

\subsection{Scaling relation between ${\cal R}^e$ and $n_t$}
Having obtained the good estimate for the average delay time $t_d$,
we next examine the relation between the active population $N_t$ and the infection rate on that day $R^e_t$.
For this analysis, we take the period $t \ge 50$ (from 5th Mar) so that the accumulated number and the daily numbers in it is enough to obtain the estimation of $R^e$ in a reasonable range: $<5.0$.
As shown in Fig. \ref{fig:fittings}, $R^e_t$ in this period obeys to an almost single-valued function of the passenger density $n_t$,
confirming the correctness of the estimated $t_d$.

The first fitting function to be tested is that of SIR-type (Lotoka-Volterra-type) model LV1:
\begin{equation}
	R^{LV1}_t = {\cal R}_0 n_t^2.
\end{equation}
This model is based on the well-mixed approximation for the collisions among people,
which in this case is consistent with an assumption that the dominant process of the infection is ``random collisions'' in the street or in some other public areas.

Keeping the SIR-type contact term, we also try a two-parameter model LV2 which assumes that there is a hidden base population $n_0$ which are active but does not appear in the passenger data:
\begin{equation}
	R^{LV2}_t = {\cal R}_0 \left( \frac{n_0 + n_t}{n_0 + 1} \right)^2.
\end{equation}

Our last model to be tested is based on an density-dependent (DD) contact rate
\begin{equation}
	C(S_t, I_t) = \frac{C_0 S_t I_t}{(S_t + I_t + R_t)^\alpha}.
	\label{eq_DDC}
\end{equation}
In comparison to the former models, this model can treat more general density-dependence and hence the more complex human activity.
For example, if one assume that the major risk of infectious contact is the intended meetings (or co-locating) with a pre-determined member size (e.g. working in the office, dining with friends, shoppings, etc), the actual contact rate should become more moderate and hence we expect positive $\alpha$.
The exponent $\alpha$ is also expected to be positive if people are trying to take a distance each other to avoid infection.
In this sense, $\alpha > 0$ can be regarded as an indication of human behavior actively perfomed. 
On the other hand, if people are in a situation that they may be trying to take a distance but sometimes they simply cannot (e.g. in crowded trains, stations, shops, etc.),
the contact rate becomes steeper and hence the exponent $\alpha$ should be negative.
This density-dependent contact rate gives the corresponding fitting function for ${\cal R}^e$:
\begin{equation}
	R^{DD}_t = {\cal R}_0 n_t^{2-\alpha}.
\end{equation}

As shown in the Table I and Fig. \ref{fig:fittings}, the density dependent contact rate model (DD) is found to give the best fit to the data.
The obtained exponent $\alpha \approx 1 > 0$ implies that the people's behavior is better characterized by their will or demand, rather than random or passive ones.
LV2 model gives the second best fit, although the obtained fitting line is similar to that of DD model.
LV1 model is the worst in the overall fitting, while it gives the best prediction in the least $n_t$ region.

The border between the region in which DD and LV2 model fit well and the region in which LV1 fit better is characterized by the sharp drop and rise of ${\cal R}^e$ at around $n_t = 0.35$. These vertical moves in the $n_t$-${\cal R}^e$ diagram correspond to the beginning and the end of the maximum alert period (from day 107 to 135: the emergency declaration was applied to the entire country).
If we take DD model or LV2 model, the origin of these jumps deviating from the theoretical fits should be the sharp change in people's behavior and closing of most shops, restaurants, etc in response to the emergency declaration.
Interestingly, similar sharp drop in ${\cal R}^e$ is observed just after the  declaration of ``Tokyo alert'' (the local alert declared by the Tokyo metropolitan gevernment), during the increase of $n_t$.

\section{Conclusions}
We have empirically investigated the relation between the effective reproduction number ${\cal R}^e_t$ the number of passengers in the main stations in Tokyo $n_t$.
The delay time from the moment of infection to the day the infection is reported is first estimated robustly at around $t_d = 15$ days, which is consistent with what is generally expected.
Based on the estimated delay time, the scaling relation between ${\cal R}^e_t$ and $n_t$ is examined.
The best fit function suggest a density-dependent correction to the rate of human contact, the exponent of which illustrates the relevance of the active aspects of human behavior.

\begin{acknowledgments}
This work was partly supported by JSPS KAKENHI grant number 18K03449TS to TS.
\end{acknowledgments}

\bibliography{ref}

\end{document}